\input amstex
\loadbold
\documentstyle{amsppt}
\magnification=\magstep1
\NoRunningHeads
\NoBlackBoxes
\TagsOnRight
\topmatter
\title   Application of Uniform Asymptotics to the Second Painlev{\'e}
Transcendent 
\endtitle
\author Andrew P. Bassom$^1$, Peter A. Clarkson$^2$, C.K. Law$^3$ and
J. Bryce McLeod$^{4,*}$ 
\endauthor 
\affil $^1$ Department of Mathematics, University of Exeter, North Park
Road, Exeter EX4 4QE, U.K.\\
$^2$ Institute of Mathematics \& Statistics, University of Kent,
Canterbury CT2 7NF, U.K.\\
$^3$ Department of Applied Mathematics, National Sun Yat-sen University,
Kaohsiung, Taiwan, R.O.C.\\
$^4$ Department of Mathematics \& Statistics, University of Pittsburgh,
Pittsburgh, PA  15260, U.S.A.
\endaffil
\thanks $^*$  The work of this author was partially supported by NSF
Grant No. DMS 95-01395, and by the Engineering and Physical Sciences
Research Council, U.K., Grant No. GR/J90985
\endthanks
\abstract In this work we propose a new method for investigating
connection problems for the class of nonlinear second-order differential
equations known as the Painlev{\'e} equations.  Such problems can be
characterized by the question as to how the asymptotic behaviours of
solutions are related as the independent variable is allowed to pass
towards infinity along different directions in the complex plane. 
Connection problems have been previously tackled by a variety of
methods.  Frequently these are based on the ideas of isomonodromic
deformation and the matching of WKB solutions.  However, the
implementation of these methods often tends to be heuristic in nature
and so the task of rigorising the process is complicated.  The method we
propose here develops uniform approximations to solutions.  This removes
the need to match solutions, is rigorous, and can lead to the solution
of connection problems with minimal computational effort.

Our method is reliant on finding uniform approximations of differential
equations of the generic form
$$\frac{{\text d}^2\phi}{{\text d}\eta^2} = - \xi^2F(\eta,\xi)\phi$$
as the complex-valued parameter $\xi \to \infty.$  The details of the
treatment rely heavily on the locations of the zeros of the function $F$
in this limit.  If they are isolated then a uniform approximation to
solutions can be derived in terms of Airy functions of suitable
argument.  On the other hand, if two of the zeros of $F$ coalesce as
$|\xi| \to \infty$ then an approximation can be derived in terms of
parabolic cylinder functions.  In this paper we discuss both cases, 
but illustrate our technique in action by applying the parabolic cylinder case
to the ``classical'' connection problem associated with the second
Painlev{\'e} transcendent.  Future papers will show how the technique can
be applied with very little change to the other Painlev{\'e} equations,
and to the wider problem of the asymptotic behaviour of the general
solution to any of these equations.
\endabstract
\endtopmatter
\document
\font\sans=cmssbx10
\def\bss#1{\hbox{\sans #1}}

\def\pain{Painlev{\'e}}
\define\bs{\boldsymbol\sigma}
\redefine\i{\text{i}}
\redefine\e{\text{e}}
\redefine\d{\text{d}}

\define\et{\eta_0} 
\define\z{\zeta}
\redefine\d{\text{d}}
\def\bS{{\bss S}}
\define\bp{\Phi}

\define\bPsi{\boldsymbol\Psi}
\define\ps1{\bPsi^{(1)}}
\define\p2{\bPsi^{(2)}}
\define\nd{D_\nu}
\define\1d{D_{-\nu-1}}
\define\ck{C_{k,k+1}}
\define\2xi{\sqrt{2\xi}}
\define\pei{\e^{\pi\i/4}}
\define\ip{\e^{-\pi\i/4}}
\def\tfr#1#2{{\textstyle{#1\over#2}}}

\subhead{1. Introduction}\endsubhead

Asymptotic behaviour of solutions of the second \pain\ transcendent (PII),
$$q'' = 2q^3+xq + \beta,\tag 1.1$$
where $' \equiv \d/\d x$ and $\beta$ is a complex constant, have been much
studied, for example in [3, 4, 6--17, 19--21, 23].
In particular connection problems have been
investigated in which one attempts to relate the asymptotic behaviour in
one $x-$ direction to that in another.  Some of the results are
heuristic, and some rigorous.  The heuristic arguments tend to use the
method of isomonodromic deformations, linked with asymptotic arguments
that use the WKB method and matching, and although Suleimanov [23] has
given a rigorous version of this for one problem associated with the
second \pain \ transcendent (1.1), the task of extending these techniques
rigorously to more complicated problems, and in particular to problems
associated with the higher equations, seems formidable.

In this paper we develop a new technique for investigating such
problems.  The technique uses the method of isomonodromy, but thereafter
develops a uniform approximation which dispenses with matching, is
rigorous and even from a computational point of view is simpler than
previous methods.  We will use it in this paper to study the asymptotic
behaviour of solutions of PII (1.1) when $\beta = 0,$ giving the
algorithm which enables one to compare asymptotic behaviour in
different directions, but we emphasise that the method is certainly not
restricted to PII, and we will return in later papers to its application
to the other transcendents.

In particular, of course, we can solve once again the ``classic''
problem for PII (1.1), which for convenience and completeness we state
here.  Its statement depends upon the following theorem, a proof of
which can be found in [8].   

\proclaim{Theorem A}  There exists a unique solution of (1.1) with
$\beta = 0$ which is asymptotic to
$a\ Ai(x)$ as $x \to + \infty, \ a$
being any positive number.  If $a < 1,$ this solution exists for all
real $x$ as $x$ decreases to $-\infty,$ and, as $x \to -\infty,$

$$q(x) \sim d|x|^{-1/4}\sin\left\{\tfr23\left|x\right|^{3/2}-
\tfr34d^2\log\left|x\right| + \gamma\right\}$$
for some constants $d, \gamma$ which depend on $a$.
\endproclaim

From this result one can easily compute more detailed asymptotics which
hold as
$x \to +\infty:$

$$\align
q(x) & = \tfr12 a \pi^{-1/2}x^{-1/4}\exp\left(-\tfr23
x^{3/2}\right)\left[1-\tfr{5}{48}x^{-3/2} + O\left(x^{-3}\right)\right],\tag
1.2a\\
r(x) & = \frac{\d q}{\d x} = -\tfr12 a \pi^{-1/2}x^{1/4}\exp\left(-\tfr23
x^{3/2}\right)\left[1 + \tfr{7}{48}x^{-3/2} +
O\left(x^{-3}\right)\right].\tag 1.2b\endalign
$$

The usual connection problem is the question of the specific dependence
of $d$ and $\gamma$ on $a$, and this is given as follows:

\proclaim{Theorem B}
$$\align
d^2 & = -\pi^{-1}\log\left(1-a^2\right),\tag 1.3a\\
\gamma & = \tfr 34 \pi - \tfr 32 d^2\log 2 - \arg \Gamma (-\tfr12
\hbox{\rm i} d^2).\tag 1.3b\endalign
$$
\endproclaim
We have already mentioned that our technique involves the concept of
isomonodromy, and we now quickly review the relevant facts [6].  (Again
we give the details for PII (1.1) but emphasise that comparable results
are known [11] for all the other \pain \ transcendents, and indeed that
there is a hierarchy of equations [1] of higher order which fit into the
same general framework.)  Suppose that $x$ and $\lambda$ are independent
complex variables and there exists a $2 \times 2$ matrix function
$\bPsi(x,\lambda)$ which satisfies both
$$\frac{\partial\bPsi}{\partial x} = (-\i\lambda \boldsymbol\sigma_3 +
q\boldsymbol\sigma_1)\bPsi, \qquad \text{i.e.}\qquad D_x\bPsi = 0,\tag 1.4$$
and
$$\frac{\partial\bPsi}{\partial\lambda} = \left\{-\i(4\lambda^2 + x +
2q^2)\bs_3 + 4\lambda q\bs_1 - 2r\bs_2 - \frac\beta\lambda\bs_1\right\}\bPsi,
\qquad \text{i.e.}\qquad D_\lambda \bPsi = 0.\tag 1.5$$
Here
$$\bs_1 = \pmatrix 0 & 1\\1 & 0\endpmatrix, \qquad \bs_2 = \pmatrix 0 &
-\i\\ \i & \ 0\endpmatrix, \qquad \bs_3 = \pmatrix 1 & 0\\ 0 &
-1\endpmatrix,$$
are the standard Pauli spin matrices which, in particular, satisfy
$$\bs_1\bs_2 = \i\bs_3, \qquad \bs_2\bs_3 = \i\bs_1, \qquad \bs_3\bs_1 =
\i\bs_2.$$
Then there is a compatibility condition
$$[D_x,D_\lambda]\bPsi = (D_xD_\lambda - D_\lambda D_x)\bPsi = 0,\tag 1.6$$
and an easy calculation shows that (1.6) reduces to (1.1).  Conversely,
if $q(x)$ evolves according to (1.1), then (1.4) and (1.5) are
compatible.  Thus (1.1) is equivalent to compatibility and compatibility
is easily seen to imply isomonodromy.

For suppose that we have two fundamental solutions $\bPsi^{(1)},
\bPsi^{(2)}$ of (1.5) in two different but overlapping sectors in the
$\lambda-$plane.  (The equation has an irregular singularity at $\lambda
= \infty$ and a regular singularity at $\lambda = 0$ but, as far as
monodromy is concerned, we need only deal with the irregular
singularity.)  Since $\bPsi^{(1)}$ and $\bPsi^{(2)}$ are both fundamental
solutions there must be a matrix $\bS$ independent of $\lambda$ but in
general dependent on $x$, such that
$$\bPsi^{(2)}(x,\lambda) = \bPsi^{(1)}(x,\lambda)\bS(x), \tag 1.7$$
and $\bS$ is referred to as the monodromy matrix.  (Of course, $\bS$ depends
on the particular fundamental solutions which are compared, and we
return to this point later.)  There is a monodromy matrix for each pair
of sectors, and the assemblage of all the monodromy matrices forms the
monodromy data.  If we now differentiate (1.7) with respect to $x$ and
use the fact that $\bPsi^{(1)}$ and $\bPsi^{(2)}$ satisfy (1.4), we
obtain immediately that $\bS$ is independent of $x$, which is to say that
the problem is isomonodromic in $x.$  It should be noted that this
involves care in choosing $\ps1$ and $\p2$, for if we multiply
$\ps1(x,\lambda)$ by a function of $x$, it still satisfies (1.5), but no
longer (1.4).

We make the remark also that we shall be able to arrange that the
monodromy matrix takes the form of a triangular matrix with 1 as the
principal diagonal.  Thus the monodromy data reduces to the one
remaining entry in the matrix, the so-called Stokes multiplier.

Given isomonodromy, we can now prove Theorem B as follows.  We work out
the monodromy data for (1.5) as $x \to + \infty,$ using the known
asymptotic dependence of $q$ on $x$, and then the monodromy data as $x
\to -\infty$, and equate them to give the required result.  The way in
which this has so far been carried out is to compute the fundamental
solutions in different sectors and use (1.7) to obtain $\bS$.  This means
that we have to compute the solutions (or at least their asymptotic
behaviours) as $|\lambda|\to \infty$ and also as $|x| \to \infty$.  This
uses WKB asymptotics, and also matching, since the form of the
asymptotics depends on the relative values of $\lambda$ and $x$, and we
have to match different forms in different regions.  The procedure can
be complicated and rigorising it difficult.

The procedure would be much simplified if one could find approximations
to solutions which are uniformly valid for all relevant large
$|\lambda|, |x|$.  This we can in fact do, and in a general form which
is certainly applicable to more than just PII (1.1).  Once it is done,
there is no further rigorous analysis required; it is merely a matter
of computing the monodromy data by relating it to the (known) data for
the approximations.

In Section 2 we describe, in the context of PII (1.1), the heuristic
reasoning which leads to the uniform approximation.  Then in Sections
3,4 we state and prove two theorems on uniform approximations, which we
believe to be the only such theorems necessary for the discussion of any
of the \pain \ equations.  In the final sections of the paper we use these
theorems to compute monodromy data both in a general setting and in the
particular case of PII, and finally as an application prove Theorem B.

It should be remarked that for the purposes of Theorem B only the first
of the two approximation theorems (that relating to double
turning-points) is required.  For more general solutions of PII, and for
a general discussion of the other \pain \ equations, the second theorem is
also required.  We intend to return to such developments in later
papers.

\subhead{2.  Deriving a Uniform Approximation}\endsubhead

To see the nature of the uniform approximation, we turn (1.5) into a
single second-order equation.  We first make the scaling
$$\xi = x^{3/2}, \qquad \eta = x^{-1/2}\lambda,$$
so that (1.5) becomes
$$\frac{\d\Psi}{\d\eta} = \xi \left\{-\i\left(4\eta^2 + 1 +
\frac{2q^2}{x}\right)\bs_3 + \left(\frac{4\eta q}{\sqrt x} -
\frac{\beta}{\eta\xi}\right) \bs_1 - \frac{2r}{x}\bs_2\right\}\Psi,$$
which, with
$$\Psi = \pmatrix \psi_1\\ \psi_2\endpmatrix,$$
is equivalent to
$$\align
\frac{\d\psi_1}{\d\eta} & = \xi \left\{-\i\left(4\eta^2 + 1 +
\frac{2q^2}{x}\right) \psi_1 + \left(\frac{4\eta q}{\sqrt x} -
\frac{\beta}{\eta \xi} + \frac{2\i r}{x}\right)\psi_2\right\},\tag 2.1a\\
\frac{\d\psi_2}{\d\eta} & = \xi\left\{\i\left(4\eta^2 + 1 +
\frac{2q^2}{x}\right)\psi_2 + \left(\frac{4\eta q}{\sqrt x} -
\frac{\beta}{\eta \xi} - \frac{2\i r}{x}\right)\psi_1\right\}.\tag
2.1b\endalign
$$    
Eliminating $\psi_1,$ we obtain
$$\align
\frac{\d^2\psi_2}{\d\eta^2} &=
\xi\left\{\i\left(4\eta^2+1+\frac{2q^2}{x}\right) \frac{\d\psi_2}{\d\eta}+ 8
\i\eta \psi_2 + \left(\frac{4q}{\sqrt x} +
\frac{\beta}{\eta^2\xi}\right)\psi_1\right.\\
&\qquad-\i\xi\left(\frac{4\eta q}{\sqrt x} - \frac{\beta}{\eta \xi} -
\frac{2\i r}{x}\right)\left(4\eta^2 + 1 + \frac{2q^2}{x}\right)\psi_1\\
&\left.\qquad+ \xi\left(\frac{4\eta q}{\sqrt x} - \frac{\beta}{\eta \xi} +
\frac{2\i r}{x}\right)\left(\frac{4\eta q}{\sqrt x} -
\frac{\beta}{\eta\xi} - \frac{2\i r}{x}\right)\psi_2\right\}\\
&=\xi\left\{-\xi\left(4\eta^2+1+\frac{2q^2}{x}\right)^2\psi_2 + 8\i
\eta\psi_2 +\xi\left[\left(\frac{4\eta q}{\sqrt x}- \frac{\beta}{\eta
\xi}\right)^2+ \frac{4r^2}{x^2}\right]\psi_2\right.\\
&\qquad+\frac{1}{\xi^2}\left(1 + \frac{\beta} {4\eta^2qx}\right)\left(\eta-\frac{\i r}
{2q\sqrt x} - \frac{\beta}{4\eta qx}\right)^{-1}\\
&\left.\qquad\times\left[\frac{\d\psi_2}{\d\eta}
- \i\xi \left(4\eta^2 + 1 +
\frac{2q^2}{x}\right)\psi_2\right]\right\}.\endalign
$$
The term $\d\psi_2/\d\eta$ can be removed by setting
$$\phi = \left(\eta - \frac{\i r}{2q\sqrt x} - \frac{\beta}{4\eta
qx}\right)^{-1/2}\psi_2,$$
whence
$$\align
\frac{\d^2\phi}{\d\eta^2} = \xi^2&\phi\left\{-(4\eta^2 + 1)^2 +
\frac{8\i\eta}{\xi} + \left(\frac{4r^2}{x^2} - \frac{4q^2}{x} -
\frac{4q^4}{x^2} - \frac{8q\beta}{x^2} + \frac{\beta^2}{\eta^2x^3}\right)\right.\\
&-\frac \i \xi \left(1 + \frac{\beta}{4\eta^2qx}\right)\left(4\eta^2 + 1 +
\frac{2q^2}{x}\right)\left(\eta-\frac{\i r}{2q\sqrt x} - \frac{\beta}{4\eta
q x}\right)^{-1}\\
&+\frac{1}{\xi^2} \frac{\beta}{4\eta^3qx}\left(\eta - \frac{\i r}{2q\sqrt
x} - \frac{\beta}{4\eta qx}\right)^{-1}\\
&\left.+\frac{3}{4\xi^2}\left(1 + \frac{\beta}{4\eta^2qx}\right)^2\left(\eta-
\frac{\i r}{2q\sqrt x} - \frac{\beta}{4\eta qx}\right)^{-2}\right\}.\tag
2.2\endalign
$$

In (2.2), attention should be drawn to the terms
$$\frac{4r^2}{x^2} - \frac{4q^2}{x} - \frac{4q^4}{x^2} -
\frac{8q\beta}{x^2} = M(\xi),\tag 2.3$$
say, which depend only on $x$ or $\xi$, and not on $\eta.$  How $M(\xi)$
behaves for large $\xi$ (which is always our interest) will depend on
the asymptotics of the functions $q(x), r(x)$ as $|x|\to \infty,$ and
therefore on the particular solution of PII.  For the remainder of this
heuristic discussion we will consider the case where $M(\xi) \to 0,$
since in the case of Theorem B this is certainly true from the given
asymptotics both as $x \to +\infty$ and as $x \to -\infty.$  But it is
not true for a general solution that $M(\xi) \to 0,$ and our methods do
not need it, and we will point out where the essential difference lies.

Assuming then that $M(\xi)\to 0$ as $|\xi| \to \infty,$ it is to be
expected from the form of (2.2) that, as $|\eta|\to\infty$ with $|\xi|$
large, the dominant term on the right-hand side will be
$$-\xi^2(4\eta^2 + 1)^2\phi,$$
so that, from the usual WKB approximation, the solution should be
asymptotically of the form
$$\eta^{-1}\exp\left\{\pm \i \xi \int^\eta (4\sigma^2 + 1)\thinspace\d\sigma\right\}
= \eta^{-1}\exp\left\{\pm \i \xi \left(\tfr43 \eta^3 +
\eta\right)\right\}.$$
The two exponentials are thus equipollent in directions
$$\arg\left(\xi \eta^3\right) = 0, \ \pm \pi, \ \pm 2\pi,\ldots,$$
i.e.
$$\arg \eta = -\tfr13 \arg \xi \pm \tfr13 k \pi, \qquad k = 0, 1, 2, \ldots$$
and these are the so-called Stokes directions.  We can determine the
Stokes multipliers by relating the asymptotic behaviour of a solution in
one Stokes direction to its asymptotic behaviour in the next, since it
is in Stokes directions (and only in Stokes directions) that the full
asymptotics appear and solutions can be defined by their asymptotics.

However, in order to connect the behaviours as $|\eta| \to \infty$ on,
say, $\arg \eta =- \frac 13 \arg \xi$ and $\arg \eta = - \frac 13 \arg
\xi + \frac 13\pi,$ we need to follow the solution along a curve for
which $\text{Re}\left\{\i \xi\int^\eta(4\sigma^2 + 1)\d\sigma\right\} = 0,$
for if we depart significantly from such curves (so-called Stokes
curves), we shall lose equipollence, and so the effect of exponentially
small solutions and therefore the Stokes multiplier.  Now there is some
choice of Stokes curve depending on the initial point of integration,
but to obtain a uniform approximation we shall consider Stokes curves
which pass through turning-points of equation (2.2); by a turning-point
we mean a value of $\eta$ which is a zero of the right-hand side of
(2.2) although we will slightly adapt this definition later.

The idea of uniform asymptotics through turning-points was first
proposed by Langer [18] and Titchmarsh [24] in work on the distribution
of eigenvalues for the Schr{\"o}dinger equation (see also [22]).  They
dealt with the equation
$$\frac{\d^2y}{\d z^2} + [\mu - q(z)]y = 0, \qquad -\infty < z <
\infty,\tag 2.4$$
where, for example, we may think of $\mu$ as a large positive parameter
and $q(z) \to \infty$ as $z \to \infty$.  If $q$ is strictly monotonic,
then there is a simple turning-point at $q(z) = \mu.$  Langer pointed
out that the prototypical case of this is $q(z) = z$, so that the
equation becomes
$$\frac{\d^2y}{\d z^2} + (\mu - z) y = 0$$
whose general solution is a linear combination of Ai($z-\mu$) and
Bi$(z-\mu),$ where Ai, Bi are the usual Airy functions.  He then went
on to show that one could obtain a uniform approximation to solutions of
(2.4), valid for large $\mu$ and $z \to \pm \infty,$ by introducing Airy
functions of a suitable argument.

We need to modify the idea further, because Langer's approximation
relates to situations where the turning-point is simple, whereas in our
case (2.2) there are two turning-points which, for large $\xi,$ are
close to $\eta = \frac 12\i$ (and two others close to $\eta = -\frac
12\i$).   (This of course is a consequence of our assumption that
$M(\xi) \to 0.$  If $M(\xi) \not\to 0,$ then the turning-points are
simple, and it is then a matter of adapting Langer's approximation using
Airy functions.)  In our present situation, therefore, it seems that the
parabolic cylinder equation 
$$\frac{\d^2y}{\d z^2} = \left[\tfr14z^2-\left(\nu + \tfr12
\right)\right]y,\tag 2.5$$
with linearly independent solutions $D_\nu(z)$ and $D_{-\nu-1}(-\i z),$ is
an appropriate one for coping with coalescing turning-points, and in
fact this possibility has already been explored by Olver [22] and
Dunster [5], primarily for real values of $z.$  With our particular
applications in mind, it will be better to consider (2.5) in the form
$$\align \frac{\d^2y}{\d z^2} & = -\xi^2 \left(z^2 - \frac{2\nu +
1}{\i\xi}\right)y\tag 2.6\\
&=-\xi^2(z^2-\alpha^2)y,\tag 2.7\endalign$$
where $\i\xi\alpha^2 = 2\nu + 1$,
which has solutions $D_\nu(\e^{\pi\i/4}\sqrt{2\xi}z)$ and
$D_{-\nu-1}(\e^{-\pi\i/4}\sqrt{2\xi}z).$

To see how this applies to (2.2), we will restrict ourselves to the
particular case when $\beta = 0.$  We try as a uniform approximation to
a solution of (2.2) the expression
$$\phi(\eta) =
\rho(\eta)D_\nu\left(\e^{\i\pi/4}\sqrt{2\xi}\zeta(\eta)\right) =
\rho(\eta)F_\nu(\zeta(\eta)), \quad\text{say},\tag 2.8$$
where the functions $\rho$ and $\zeta$ are to be determined. 
Substituting (2.8) in (2.2) with $\beta = 0$ we have
$$\align
\rho''F_\nu & + 2\rho'\zeta'F'_\nu + \rho((\zeta')^2F''_\nu +
\zeta''F'_\nu)\\
&=\xi^2\rho F_\nu\left\{-(4\eta^2+1)^2 + \frac{8\i\eta}{\xi} +
\left(\frac{4r^2}{x^2} - \frac{4q^2}{x} - \frac{4q^4}{x^2}\right)\right.\\
&\left.-\frac\i \xi \left(4\eta^2 + 1 + \frac{2q^2}{x}\right)\left(\eta-\frac{\i
r}{2q\sqrt x}\right)^{-1} + \frac{3}{4\xi^2}\left(\eta -\frac{\i r}{2q\sqrt
x}\right)^{-2}\right\}.\tag 2.9\endalign
$$
Recalling that $F_\nu$ satisfies (2.6) we can compare coefficients of
$F'_\nu$ and $F_\nu$ in (2.9).  The vanishing of the coefficient of
$F'_\nu$ gives
$$2\rho'\zeta' + \rho \zeta'' = 0,$$
so that we can take
$$\rho = (\zeta')^{-1/2},\tag 2.10$$
for the choice of integration constant at this point is inconsequential.
 The vanishing of the coefficient of $F_\nu$ gives
$$\align
\xi^2&(\zeta^2-\alpha^2)(\zeta')^2\\
&=\xi^2\left\{(4\eta^2 + 1)^2 - \frac{8\i \eta}{\xi} -
\left(\frac{4r^2}{x^2} - \frac{4q^2}{x} - \frac{4q^4}{x^2}\right)\right.\\
&\left.+\frac \i \xi\left(4\eta^2 + 1 + \frac{2q^2}{x}\right)\left(\eta - \frac{\i
r}{2q\sqrt x}\right)^{-1} - \frac{3}{4\xi^2}\left(\eta - \frac{\i
r}{2q\sqrt x}\right)^{-2} + \frac{\rho''}{\xi^2\rho}\right\}.\tag
2.11\endalign
$$
If we ignore the last two terms in $\{\ldots\}$ as being of smaller
order (for large $\xi$) than the others, then we are left with
$$\align
(\zeta^2-\alpha^2 )(\zeta')^2 &= (4\eta^2 + 1)^2 - \frac{8\i \eta}{\xi}
-\left(\frac{4r^2}{x^2} - \frac{4q^2}{x} - \frac{4q^4}{x^2}\right)\\
&\qquad + \frac \i \xi \left(4\eta^2 + 1 + \frac{2q^2}{x}\right)\left(\eta -
\frac{\i r}{2q\sqrt x}\right)^{-1}\\
&=G(\eta,\xi), \quad \text{say,} \tag 2.12\endalign
$$
which, apart from a constant of integration, defines $\zeta$ as a
function of $\eta$ once we have specified $\alpha.$  (We recall always
that $r, q, x, \xi$ are constants as far as $\eta, \zeta$ are
concerned.)  We note however from (2.10) that we certainly want to avoid
zeros of $\zeta'$ and, from (2.12), $\zeta'$ has a zero wherever
$G(\eta,\xi) = 0;$ i.e. essentially at a turning-point of the equation,
unless we can choose $\alpha$ so that the zeros of $\zeta^2-\alpha^2$
coincide with those of $G$.  For large $\xi,$ there are two
turning-points, say $\eta_1$ and $\eta_2$, close to $\frac 12\i,$ and
two close to $-\frac 12\i.$  If we are interested in a Stokes curve
which passes through (or close to) $\frac12\i$, then we must choose
$\alpha$ so that $\zeta = -\alpha$ corresponds to $\eta = \eta_1$ and
$\zeta = +\alpha$ corresponds to $\eta = \eta_2.$  We can ensure one of
these holds by use of the constant of integration implicit in the
evaluation of $\zeta$ from (2.12).  The second can be achieved by
defining $\alpha$ so that
$$\int^\alpha_{-\alpha}(\zeta^2-\alpha^2)^{1/2}\thinspace\d\zeta =
\int^{\eta_2}_{\eta_1}G^{1/2}(\eta,\xi) \thinspace\d\eta.$$
Since the left-hand side integrates easily to $\frac 12\pi \i\alpha^2,$
we have $\alpha$ given by
$$\tfr12\pi \i \alpha^2 =
\int^{\eta_2}_{\eta_1}G^{1/2}(\eta,\xi)\thinspace\d\eta.\tag 2.13$$
With $\alpha$ so defined, and $\zeta$ chosen according to
$$\int^\zeta_\alpha(\tau^2-\alpha^2)^{1/2}\thinspace\d\tau =
\int^\eta_{\eta_2}G^{1/2}(\sigma,\xi)\thinspace\d\sigma,$$
we can hope that solutions of (2.2) are approximated, uniformly on
$\eta$ for large $\xi,$ by some linear combination of
$$(\zeta')^{-1/2}D_\nu(\e^{\i\pi/4}\sqrt{2\xi}\zeta)\quad\text{and}\quad
(\zeta')^{-1/2}D_{-\nu-1}(\e^{-\i\pi/4}\sqrt{2\xi}\zeta).$$
A precise statement and proof of this conjecture is given in the next
section.

We remark finally that it is a consequence of this uniform approximation
that the monodromy data for (2.2) as $|\xi| \to \infty$ will be the same
as that for the parabolic cylinder functions, which can be found in any
text on special functions, modified only by some allowance for the
various changes of variable involved.  We work this out more precisely
in Sections 5 \& 6.

\subhead{3.  The Uniform Approximation Theorem for a Double
Turning-Point}\endsubhead

We are interested in differential equations of the form
$$\frac{\d^2\phi}{\d\eta^2} =-\xi^2F(\eta,\xi)\phi\tag 3.1$$
and, guided by the heuristic discussion in Section 2, we shall make the
following assumptions about $F$.  Suppose that our concern is with the
limit $|\xi|\to\infty$ with $\arg \xi \to \theta;$ we then hypothesise
that
\proclaim{H1}  There is a sequence of values $\xi_n, |\xi_n|\to \infty,
\ \arg\xi_n\to\theta,$ such that
$$F(\eta,\xi_n) = F_0(\eta)\left(\eta-\eta_0\right)^2 - \frac{\widetilde
F(\eta,\xi_n)}{\xi_n},$$ where
\itemitem{(i)}  $F_0(\eta)$ is a polynomial in $\eta,$ with $F_0(\eta_0)
\ne 0$ and
$$F_0(\eta) \sim \text{A}\eta^m \quad\text{as}\quad  \eta \to \infty, \tag 3.2$$
\itemitem{(ii)}  $\widetilde F(\eta,\xi_n)$ is a rational function of
$\eta,$ with the location of its poles possibly dependent on $\xi_n.$

(Further assumptions on $\widetilde F$ are given in due course.)
\endproclaim

\remark{Remarks}  
\item{1.}  The assumption that $F_0$ is polynomial is not essential. 
Polynomial-like behaviour of some sort would certainly be sufficient,
but in applications to the \pain \ transcendents it is always the case
that $F_0$ is a polynomial, and since no new ideas would be involved in
generalization, we do not consider this here.  Similar comments hold
with regard to the rational behaviour of $\widetilde F.$
\item{2.}  The assumption $F_0(\eta_0) \ne 0$ is crucial.  It implies
that (3.1) is to have a double turning-point at $\eta = \eta_0$ (or,
more precisely, for large $|\xi|,$ two turning-points close to
$\eta_0$), but no other turning-points close to $\et.$
\item{3.}  Our assumption is only about a sequence of values $\xi_n$
since it will turn out in our applications to be a consequence of
isomonodromy that behaviour as $|\xi|\to \infty$ through a sequence is
sufficient to determine behaviour as $|\xi|\to\infty$ generally. 
However, in the present approximation theorem, which is in itself quite
independent of the concept of isomonodromy, we will not be involved in
comparing different sequences, and so we can without confusion drop the
subscript in $\xi_n,$ and this will be done henceforth.
\item{4.}  The usual WKB approximation for (3.1) would suggest, in view
of (3.2), that, for large $\eta,$ the solutions of (3.1) are asymptotic
to linear combinations of
$$\eta^{-(m+2)/4}\exp\left(\pm\i\xi \int^\eta
F^{1/2}_0(s)\left(s-\et\right)\thinspace\d s\right),$$
and so for Stokes directions we must have
$$\arg\left(\xi A^{1/2}\eta^{(4+m)/2)}\right)= 0, \ \pm \pi, \ \pm 2\pi, \ldots$$
or
$$\left(\tfr12m+2\right)\arg\eta = -\arg \xi - \tfr12\arg A + k\pi
\qquad (k = 0, \pm 1, \ldots).\tag 3.3$$
To compute monodromy data for (3.1), we need the behaviour of solutions
in two successive Stokes directions, and this leads to the next
hypothesis.
\endremark

\proclaim{H2}  There exists a Stokes curve $C_{k,k+1}$, defined by
$$\hbox{\rm Re}\left(\hbox{\rm i}\xi \int^\eta_{\et}F^{1/2}(\sigma, \xi)
\thinspace\hbox{\rm d}\sigma\right) =0,$$
which connects $\infty$ in two successive Stokes directions (given by
$k\pi$ and $(k + 1)\pi$ in (3.3) above),  passes through $\et$
and is (for large $\xi$) bounded from any zero of $F_0.$
\endproclaim

\proclaim{H3}  On and in a neighbourhood of $C_{k,k+1}, \widetilde F$
has no poles, at least for large $\xi,$ and, for all $\eta$ and uniformly
for large $\xi,$
$$\frac{\widetilde F(\eta,\xi)}{F_0(\eta)} = O (|\eta| + 1),$$
while, for large $\eta$ and uniformly for large $\xi$,
$$F'/F = O (\eta^{-1}), \qquad F''/F = O (\eta^{-2}), \qquad F' =
dF/d\eta.$$
\endproclaim

\remark{Remarks}  
\item{1.}  It is now clear from Rouch{\'e}'s theorem that, for $\xi$
sufficiently large, $F(\eta,\xi)$ has, as a function of $\eta,$
precisely two zeros $\eta_1,\eta_2$ close to $\eta_0.$  In fact, if
$\widetilde F(\eta,\xi)\to L$ as $\eta\to\et, |\xi| \to \infty,$ we have
$$\eta_j = \et \pm\left(\frac{L}{F_0(\et)}\right)^{1/2}
\xi^{-1/2}\{1+o(1)\}, \qquad j = 1,2.\tag 3.4$$
(We can take $\eta_2$ to correspond to the upper sign.)
\item{2.}  In line with the heuristic discussion in Section 2, we define
a number $\alpha$ by
$$\tfr12\pi\i\alpha^2 = \int^\alpha_{-\alpha} 
(\tau^2-\alpha^2)^{1/2}\thinspace\d\tau =
\int^{\eta_2}_{\eta_1}F^{1/2}(\eta,\xi)\thinspace\d\eta,\tag 3.5$$
and a new variable $\zeta$ by
$$\int^\z_\alpha(\tau^2-\alpha^2)^{1/2}\thinspace\d\tau =
\int^\eta_{\eta_2}F^{1/2}(s,\xi)\thinspace\d s.\tag 3.6$$
There is a choice of signs for the various square roots, but any
consistent choice will do.  Other choices merely lead to a permutation
amongst the solutions $D_\nu(z), D_\nu(-z), D_{-\nu-1}(\i z)$ and
$D_{-\nu-1}(-\i z)$ of (2.5) (or, of course, (2.6)) and do not therefore
affect the space of approximating functions in our theorem below.  We
note also that $F$ does not vanish on or near $C_{k,k+1}$ if $\xi$ is
large, except at $\eta_1, \eta_2,$ and so there is no ambiguity in the
sign of $F^{1/2}$ once some initial value has been chosen.
\item{3.}  There is a certain arbitrariness in the precise choice of a
Stokes curve.  All that is required is that on it both WKB
approximations are equipollent, so that both appear in an asymptotic
expansion of a solution.  With this in mind, it would be equally good to
choose a curve connecting two Stokes directions on which
$\text{Re}(\i\xi\int^\eta_{\et} F^{1/2}(\sigma,\xi)\thinspace
\d\sigma)$ is bounded
independently of $\eta$ and $\xi,$ and we shall make use of this
possibility.

Given these three assumptions concerning the problem (3.1) we can now
show that solutions of this equation can be approximated uniformly by
parabolic cylinder functions so long as $\eta$ remains on $C_{k,k+1}.$ 
This result can be summarised thus:
\endremark

\proclaim{Theorem 1}  Under hypotheses H1-H3, and given any solution
$\phi$ of (3.1), there exist constants $c_1$, $c_2$ such that, uniformly
for $\eta$ on $C_{k,k+1},$ as $|\xi|\to\infty,$
$$\align
\left(\frac{\z^2-\alpha^2}{F(\eta,\xi)}\right)^{-1/4}\phi(\eta,\xi)
&=\left\{[c_1+o(1)]D_{\nu}\left(\e^{\pi\i/4}\sqrt{2\xi}\zeta\right)\right.\\
&\left.\qquad\qquad+[c_2+o(1)]D_{-\nu-1}\left(\e^{-\pi\i/4}\sqrt{2\xi}\z\right)\right\}.\endalign$$
\endproclaim

\demo{Proof}  We have to compare the equations
$$\frac{\d^2\phi}{\d\eta^2} = -\xi^2F(\eta,\xi)\phi\tag 3.7$$
and
$$\frac{\d^2\psi}{\d\z^2} = -\xi^2(\z^2-\alpha^2)\psi.\tag 3.8$$
Set
$$p=\frac{\d\eta}{\d\z} = \left(\frac{\z^2-\alpha^2}{F}\right)^{1/2},\tag 3.9$$
which we note is bounded both above and below on any bounded part of
$C_{k,k+1}.$  The only problem can occur near $\et$, and there we notice
that $\z^2-\alpha^2$ and $F$ have the same zeros, so that  $p$ and $p^{-1}$
are analytic in a neighbourhood of $\et$.  Since trivially $p$ and
$p^{-1}$ are bounded on some fixed small circle $|\eta-\et|=k,$ say, it
follows from the maximum principle that $p$ and $p^{-1}$ are bounded
inside $|\eta-\et|=k.$  Also as $\eta, \z\to\infty,$ it is immediate from
(3.6) that
$$\tfr12\z^2\sim \frac{2A^{1/2}\eta^{2+m/2}}{4+m},$$
so that $p$ is asymptotically some power of $\eta$ (or $\z$), and so,
considering $p=p(\z)$ and $p' = dp/d\z,$ we have, as $|\z|\to\infty,$
from H3,
$$\frac{p'}{p} = O \left(\frac 1\z\right),\quad\quad \frac{p''}{p} = O\left(
\frac {1}{\z^2}\right),$$
and the bounds implicit in the $O-$terms are independent of $\xi.$
Now
$$\frac{\d\phi}{\d\eta} = \frac 1p \frac{\d\phi}{\d\z}, \quad
\frac{\d^2\phi}{\d\eta^2} =
\frac{1}{p^2}\frac{\d^2\phi}{\d\z^2}-\frac{p'}{p^3}
\frac{\d\phi}{\d\z},$$
so that (3.7) becomes
$$\frac{\d^2\phi}{\d\z^2} = -\xi^2(\z^2-\alpha^2)\phi +
\frac{p'}{p}\frac{\d\phi}{\d\z}.$$
Setting
$$\phi=p^{1/2}\Phi,\tag 3.10$$
we have
$$\frac{\d^2\Phi}{\d\z^2} = -\xi^2(\z^2-\alpha^2)\bp-\frac
12\left[\frac{p''}{p}-\frac 32\frac{(p')^2}{p^2}\right]\bp.\tag 3.11$$
Now we have already seen that linearly independent solutions of (3.8)
are
$$D_\nu\left(\e^{\pi\i/4}\sqrt{2\xi}\z\right),\qquad
\1d\left(\e^{-\pi\i/4}\sqrt{2\xi}\z\right),\tag 3.12$$
where, by (2.7),
$$\nu = -\tfr12 + \tfr12\i\xi\alpha^2,\tag 3.13$$
and the asymptotics of the functions in (3.12), as $|\sqrt{2\xi}\z| \to
\infty,$ are always linear combinations of
$$\exp\left(-\tfr12
\i\xi\z^2\right)\left(\sqrt{2\xi}\z\right)^\nu\quad\text{and}\quad
\exp\left(\tfr12\i\xi\z^2\right)\left(\sqrt{2\xi}\z\right)^{-\nu-1}.$$
(For the asymptotics of parabolic cylinder functions, one can consult,
for example, [25].)  We want to assert that these are bounded on
$C_{k,k+1}$, which is so if
$$\text{Re}\left(\tfr12\i\xi\z^2-\nu\log\left(\sqrt{2\xi}\z\right)\right)
\qquad \text{is bounded}.\tag 3.14$$
But by definition
$$\i\xi\int\limits^\zeta_\alpha(\tau^2-\alpha^2)^{1/2}\d\tau =
\i\xi\int\limits^\eta_{\eta_2}F^{1/2}(s,\xi)\d
s=\i\xi\int\limits^\eta_{\et}F^{1/2}(s,\xi)\d s + \i
\xi\int\limits^{\et}_{\eta_2}F^{1/2}(s, \xi) \d s,$$
and the last term is bounded independent of $\xi.$  (Merely set $s-\et =
t\xi^{-1/2}$ in the integrand, and use the fact that
$(\eta_2-\et)\xi^{1/2}$ is bounded.)  Thus, on $\ck,$
$$\text{Re}\left(\i\xi\int\limits^\z_\alpha (\tau^2-\alpha^2)^{1/2}\d
\tau\right)\quad\text{is bounded,}\tag 3.15$$
and it is an elementary integration that
$$\int\limits^\z_\alpha(\tau^2-\alpha^2)^{1/2}\d\tau = \tfr12
\left\{\z(\z^2-\alpha^2)^{1/2}-\alpha^2\log\left(\z +
(\z^2-\alpha^2)^{1/2}\right) + \alpha^2\log\alpha\right\}.\tag 3.16$$
Substituting for $\alpha$ from (3.13), we see easily that (3.15) implies
(3.14).

We can now turn (3.11) into an integral equation in the usual way.  In
fact, any solution of (3.11) satisfies, for some constants $c_1,c_2,$ the
integral equation
$$\align
\bp(\zeta)&=c_1\nd\left(\e^{\pi\i/4}\sqrt{2\xi}\z\right)+c_2\1d
\left(\e^{-\pi\i/4}\sqrt{2\xi}\z\right)\\
&-\frac{\i}{2\sqrt{2\xi}}\int\limits^\z_\alpha\left\{\nd\left(\e^{\pi\i/4}
\sqrt{2\xi}\z\right)\1d\left(\e^{-\pi\i/4}\sqrt{2\xi}t\right)\right.\\
&\left.\qquad-\1d\left(\e^{\pi\i/4}\sqrt{2\xi}\z\right)\nd\left(\e^{-\pi\i/4}
\sqrt{2\xi}t\right)\right\}\left[\frac{p''}{p}-\frac
32 \frac{(p')^{2}}{p^2}\right]\bp(t)\thinspace \d t.\tag 3.17\endalign
$$
In deriving (3.17) we have made use of the standard result that the Wronskian
$$W\left(\nd\left(\pei\2xi \z\right), \1d\left(\ip\2xi \z\right)\right) =
\i\2xi$$
and the integral is to be taken along $\ck.$  Since $\nd, \1d$ are
bounded on this curve, and
$$\frac{p''}{p} - \frac 32 \frac{(p')^{2}}{p^2} = O
\left(\frac{1}{\z^2}\right)$$
and so is integrable to infinity on $\ck$, we can solve (3.17) by
iteration (see, for example [24], to conclude that $\bp$ is bounded on
$\ck$.  Furthermore, we deduce that
$$\bp(\z) = c_1\nd\left(\pei\2xi\z\right) + c_2\1d\left(\ip\2xi\z\right) + O
\left(\frac{|c_1| + |c_2|}{\sqrt\xi}\right)$$
and, returning to $\phi$ via the transformation (3.10), we see that the
theorem is proved.
\enddemo
\subhead{4.  The Uniform Approximation Theorem for a Simple
Turning-Point}\endsubhead

Consider differential equations of the form
$$\frac{\d^2\phi}{\d\eta^2} = - \xi^2F(\eta,\xi) \phi,\tag 4.1$$
where we make the following assumptions about $F$ in the limit as
$|\xi|\to\infty, \arg \xi = \theta.$
\proclaim{H1}  There is a sequence of values $\xi_n, |\xi_n|\to\infty,
\arg\xi_n\to \theta,$ such that
$$F(\eta,\xi_n) = F_0(\eta,\xi_n)\left(\eta-\eta_0(\xi_n)\right)- \frac{\widetilde
F(\eta,\xi_n)}{\xi_n},\tag 4.2$$
where
\itemitem{(i)}  $\eta_0(\xi_n) \to \eta_\infty$ as $\xi_n\to\infty,
\eta_\infty$ finite,
\itemitem{(ii)}  $F_0(\eta,\xi_n)$ is a polynomial in $\eta$ whose zeros
tend to finite limits as $\xi_n\to\infty,$ all distinct from
$\eta_\infty,$ and
$$F_0(\eta,\xi_n) \sim A\eta^m \quad\text{as}\quad \eta \to \infty, \tag 4.3$$
\item{(iii)}  $\widetilde F(\eta,\xi_n)$ is a rational function of
$\eta.$
\endproclaim

\remark{Remarks}
\item{1.}  We are allowing the possibility that the turning-point $\et$
may depend on $\xi$.  (We drop the subscript $n$ as in Section 3.)  We
could do this also in Theorem 1, but this does not seem relevant in the
applications of Theorem 1, whereas it certainly is in applications of
the present case.
\item{2.}  The usual WKB approximation for (4.1) would suggest, from
(4.3), that for large $\eta$ solutions of (4.1) are asymptotic to linear
combinations of
$$\eta^{-(m+1)/4}\exp\left(\pm\i\xi\int^\eta
F^{1/2}_0(s,\xi)(s-\et)^{1/2}\d s\right)$$
and so for Stokes directions we must have
$$\arg\left(\xi A^{1/2}\eta^{(3+m)/2}\right) = 0,\quad \pm \pi, \quad \pm 2\pi,
\ldots$$
or
$$\tfr12(m+3)\arg \eta =-\arg \xi - \tfr12  \arg A + k\pi \qquad (k
= 0, \pm 1, \ldots).\tag 4.4
$$

Monodromy data for (4.1) can be computed once the behaviours of
solutions in two successive Stokes directions are known.  We therefore
assume the following:

\proclaim{H2}  There exists a Stokes curve $\ck,$ defined by
$$\hbox{\rm Re}\left(\hbox{\rm i}\xi \int^\eta_{\et}F^{1/2}(\sigma,\xi) 
\hbox{\rm d}\sigma\right) = 0,$$
which connects $\infty$ in two successive Stokes directions (given by
$k\pi$ and $(k+1)\pi$ in (4.4) above) and which passes through $\et$ and
is (for large $\xi$) bounded from any zero of $F_0.$  (We drop the
explicit dependence of $\et$ on $\xi.$)
\endproclaim

\proclaim{H3}  On and in a neighbourhood of $\ck, \widetilde F$ has no
poles, at least for large $\xi,$ and, for all $\eta$ and uniformly for
large $\xi,$
$$\frac{\widetilde F(\eta,\xi)}{F_0(\eta,\xi)} = O(1),$$
whilst, for large $\eta$ and uniformly for large $\xi,$
$$F'/F = O(\eta^{-1}),\qquad F''/F = O(\eta^{-2}).$$
\endproclaim

Based on the above, it follows from Rouch{\'e}'s theorem that, for
sufficiently large $\xi, F(\eta,\xi)$ has, as a function of $\eta,$
precisely one zero $\eta^*$ close to $\et$ and so close to
$\eta_\infty.$  In fact,  
$$\eta^* = \et + O(\xi^{-1}).\tag 4.5$$

Now if we define a new variable $\z$ by
$$\tfr23\z^{3/2} = \int^\z_0\tau^{1/2}\d\tau =
\int^\eta_{\eta^*}F^{1/2}(s,\xi)\d s,\tag 4.6$$
then we can obtain uniform approximations to the solution of (4.1)
according to

\proclaim{Theorem 2}  Under hypotheses H1-H3, and given any solution
$\phi$ of (4.1), there exist constants $c_1,c_2$ such that, uniformly
for $\eta$ on $\ck,$ as $|\xi| \to \infty,$
$$\left(\frac{\z}{F(\eta,\xi)}\right)^{-1/4}\phi(\eta,\xi) =
\left\{[c_1+o(1)]\hbox{\rm Ai}
\left(\hbox{$\hbox{\rm e}^{\pi{\i}/3}$}\xi^{{2}/{3}}z\right) + [c_2 +
o(1)]\hbox{\rm Bi}
\left(\hbox{$\hbox{\rm e}^{\pi{\i}/3}$}\xi^{{2}/{3}}\z\right)\right\}$$ where $\hbox{\rm Ai}$
and $\hbox{\rm Bi}$ are the usual Airy functions.
\endproclaim

\demo{Proof}  We need to compare the equations
$$\frac{\d^2\phi}{\d\eta^2} = -
\xi^2F(\eta,\xi)\phi\qquad\text{and}\qquad
\frac{\d^2\psi}{\d\z^2}=-\xi^2\z\psi,\tag 4.7a,b$$
and do so by setting
$$p = \frac{\d\eta}{\d\z} = \left(\frac \z F\right)^{1/2}.$$
Now $p$ is bounded both above and below on any bounded part of $\ck.$ 
The only difficulty might arise near $\et$ and there we note that $\z$
and $F$ have the same simple zero (from definition (4.6)) so that $p$
and $p^{-1}$ are analytic near $\et$.  Since trivially $p$ and $p^{-1}$
are bounded on some fixed small circle $|\eta-\et| = \varepsilon,$ say, it
is a consequence of the maximum principle that both $p$ and $p^{-1}$ are
bounded inside $|\eta-\et| = \varepsilon.$  As $\eta,\z\to\infty,$ it is
obvious from (4.6) that
$$\tfr23\z^{3/2} \sim \frac{2A^{1/2}\eta^{(m+3)/2}}{m+3},\tag
4.8$$
so that $p$ is asymptotically some power of $\eta$ (or $\z$). 
Therefore, considering $p = p(\z)$ and $p'\equiv \d p/\d\z$ we have as
$|\z| \to \infty,$ from H3,
$$\frac {p'}{p} = O\left(\frac 1\z\right), \quad   \frac{p''}{p} =
O\left(\frac {1}{\z^2}\right),$$
and the bounds implicit in the $O$-terms are independent of $\xi.$
Since
$$\frac {\d\phi}{\d\eta} = \frac 1p \frac{\d\phi}{\d\z}, \quad
\frac{\d^2\phi}{\d\eta^2} = \frac{1}{p^2} \frac{\d^2\phi}{\d\z^2} -
\frac{p'}{p^3} \frac{\d\phi}{\d\z},$$
equation (4.7a) becomes
$$\frac{\d^2\phi}{\d\z^2} = -\xi^2\z\phi + \frac{p'}{p}
\frac{\d\phi}{\d\z},$$
and on setting
$$\phi = p^{1/2}\bp\tag 4.9$$
we obtain
$$\frac{\d^2\bp}{\d\z^2} = -\xi^2\z\bp - \frac 12 \left[\frac{p''}{p} -
\frac 32 \frac{(p')^2}{p^2}\right]\bp.\tag 4.10$$
It is a standard result that linearly independent solutions of (4.7b)
are
$$\text{Ai}\left(\e^{\i\pi/3}\xi^{2/3}\z\right), \qquad
\text{Bi}\left(\e^{\i\pi/3}\xi^{2/3}\z\right)\tag 4.11$$ 
and the asymptotics of these functions as $|\xi^{2/3}\z|\to \infty$ are
always linear combinations of
$$\xi^{-1/6}\z^{-1/4}\exp\left\{\pm \tfr23\i\xi\z^{3/2}\right\}.$$

We would like to assert that these are bounded on $\ck$, which is the
case if $\text{Re}\left(\i\xi\z^{3/2}\right)$ is bounded.  However, we
have from (4.6) that
$$\tfr23\i\xi\z^{3/2} = \i\xi\int^\eta_{\eta^*}F^\frac
12(s,\xi)\thinspace\d s = \i\xi \int^{\et}_{\eta^*} F^{1/2}
(s,\xi)\thinspace\d s +
\i\xi\int^\eta_{\et} F^{1/2}(s,\xi)\thinspace\d s$$
so that, on $\ck, \text{Re}(\i\xi\z^{3/2})$ is bounded if and only
if $\text{Re}\left(\i\xi\int^{\et}_{\eta^*} F^{1/2}(s,\xi)\d s\right)$ is
bounded.  This latter expression is
$O\left(|\xi|^{-1/2}\right)$ for large $|\xi|$ (using (4.2) and (4.5))
so that $\text{Re}\left(\i\xi\z^{3/2}\right)$ is indeed bounded on
the Stokes curve.

To complete the proof of Theorem 2 we turn (4.10) into an integral
equation in the usual way.  It follows that any solution of (4.10)
satisfies, for some constants $c_1$ and $c_2,$ the equation
$$\align
\bp(\z)  = c_1\text{Ai}&\left(\e^{\pi\i/3}\xi^{2/3}\z\right) +
c_2\text{Bi}\left(\e^{\pi\i/3}\xi^{2/3}\z\right)\\
&-\frac{\i}{4\xi^{5/6}}\int^\z_0\left\{\text{Ai}\left(\e^{\pi\i/3}
\xi^{2/3}\z\right)\text{Bi}\left(\e^{\pi\i/3}\xi^{2/3}t\right)\right.\\
&\left.\qquad-\text{Bi}\left(\e^{\pi\i/3}\xi^{2/3}\z\right)
\text{Ai}\left(\e^{\pi\i/3}\xi^{2/3}t\right)\right\}\left[\frac{p''}{p}-\frac
32\left(\frac{p'}{p}\right)^2\right]\bp(t)\thinspace\d t.\tag 4.12\endalign$$ In
deriving (4.12) we have used the standard result for Wronskians that
$$W\left(\text{Ai}\left(\e^{\pi\i/3}\xi^{2/3}\z\right),
\text{Bi}\left(\e^{\pi\i/3}\xi^{2/3}\z\right)\right) = 2\i\xi^{5/6}.$$
The integral within (4.12) is taken along the Stokes curve $\ck$ and
since Ai and Bi are bounded there and
$$\frac{p''}{p} - \frac 32\left(\frac{p'}{p}\right)^2 =
O\left(\frac{1}{\z^2}\right)$$
and so is integrable to infinity on $\ck$, we can solve (4.12) by
iteration to conclude that $\bp$ is bounded on $\ck$.  Furthermore, we
have that
$$\bp(\z) = c_1\text{Ai}\left(\e^{\pi\i/3}\xi^{2/3}\z\right) +
c_2\text{Bi}\left(\e^{\pi\i/3}\xi^{2/3}\z\right) + O\left(\frac{|c_1| +
|c_2|}{\xi^{5/6}}\right)$$ and, returning to the variable $\phi$ via the
transformation (4.9), we conclude that the theorem is proved.
\enddemo

\subhead{5. Monodromy Data for Parabolic Cylinder Functions}\endsubhead

This section sets out the well-known results that we shall need
concerning Stokes multipliers for the parabolic cylinder function.  We
shall be interested in computing the multipliers for the curve $\ck;$
i.e. we wish to compare the asymptotic behaviours on
$$\left(\tfr12m + 2\right)\arg\eta + \arg \xi + \tfr12\arg A = \quad
k\pi \quad\text{and}\quad (k+1)\pi$$
and, since for large $\eta,\z, 2 \arg \z\sim\frac 12\arg A + (\frac
12m+2)\arg \eta,$ this is equivalent to comparing behaviours on
$$\arg\left(\2xi\z\right) = \tfr12 k\pi \quad\text{and}\quad \tfr12(k +
1)\pi.$$
Let us set $z \equiv \e^{\pi\i/4}\2xi\z;$ the complete asymptotic
behaviours of $\nd(z)$ as $|z|\to\infty$ are well known (see for example
[2]) and are given by
$$\nd(z)\sim\cases z^\nu\exp(-\frac 14z^2),\quad & \text{if} \ |\arg z|
< \frac 34\pi,\\
{\displaystyle z^\nu\exp(-\tfr
14z^2)-\frac{\sqrt{2\pi}}{\Gamma(-\nu)}\e^{\i\pi\nu}z^{-\nu-1}\exp(\tfr 14z^2)},
\quad & \text{on} \ \arg z = \frac 34\pi,\\
{\displaystyle\e^{-2\i\pi\nu}z^\nu\exp(-\tfr 14 z^2)-\frac{\sqrt{2\pi}}{\Gamma(-\nu)}
\e^{\i\pi\nu}z^{-\nu-1}\exp(\tfr 14z^2)},  & \text{on} \ \arg
z=\frac 54\pi,\\
{\displaystyle\e^{-2\i\pi\nu}z^\nu\exp(-\tfr 14z^2)}, \ & \text{if} \ \frac 54\pi
< \arg z < \frac{11}{4}\pi. \endcases\tag 5.1$$

Then, on $\arg z = \pm \frac 14\pi + 2\ell\pi,$ with $\ell$ integral,
$$\nd(z\e^{-2\i\ell\pi})\sim(z \e^{-2\i\ell\pi})^\nu\exp\left(-\tfr14
z^2\right)$$ and so, since $\nd(z)$ is single-valued,
$$\nd(z) \sim \exp \left(-\tfr14z^2\right)z^\nu \e^{-2\ell\pi\i\nu}.\tag
5.2$$
Similarly, on $\arg z = \frac 34\pi + 2\ell\pi,$ we have
$$\nd(z) \sim \exp\left(-\tfr14z^2\right)z^\nu \e^{-2\ell\pi\i\nu} -
\frac{\sqrt{2\pi}}{\Gamma(-\nu)}\e^{\nu\pi\i}\exp\left(\tfr14
z^2\right)z^{-\nu-1} \e^{2\ell\pi\i(\nu + 1)}.\tag 5.3$$

To evaluate the Stokes multiplier we proceed as follows.  In any sector
between two adjacent Stokes directions there is (modulo multiplication
by a constant) a unique solution $f_1$ which is asymptotic to the small
exponential.  All other solutions are necessarily asymptotic to some
multiple of the large exponential, but if we take such a solution on the
first Stokes line, then we will find that on the second Stokes line its
asymptotics will have added a multiple of $f_1.$  That multiple is the
Stokes multiplier.  Thus, relative to the asymptotic forms $\exp(-\frac
14z^2)z^\nu$ and $\exp(\frac 14z^2)z^{-\nu-1},$ the Stokes multiplier
for the sector $\frac 14\pi + 2\ell\pi$ to $\frac 34\pi + 2\ell\pi,$ in
which $\exp(-\frac 14z^2)z^\nu$ is dominant, can be immediately deduced
from (5.2) and (5.3).  Consequently,
$$\text{SM}\left(\tfr14\pi + 2\ell\pi, \tfr34\pi + 2\ell\pi\right) = -
\frac{\sqrt{2\pi}}{\Gamma(-\nu)} \,\e^{\nu\pi\i}\e^{4\ell\pi\i\nu}.\tag 5.4$$
Similar calculations for each of the other pairs of sectors yields the
complete monodromy data in the form
$$\align
\text{SM}\left(\tfr34\pi + 2\ell\pi, \tfr54\pi + 2\ell\pi\right) & =
\frac{\Gamma(-\nu)}{\sqrt{2\pi}} \,\e^{-\nu\pi\i}\e^{-4\ell\pi\i\nu}(1 -
\e^{-2\pi\i\nu})\\
&=\i\sqrt{\frac{2}{\pi}}\,\Gamma(-\nu)\e^{-(4\ell+2)\pi\i\nu}\sin\pi\nu,\tag
5.5\\
\text{SM}\left(-\tfr34\pi + 2\ell\pi, - \tfr14\pi +
2\ell\pi\right)&=\frac{\sqrt{2\pi}}{\Gamma(-\nu)}\,
\e^{-\nu\pi\i}\e^{4\ell\pi\i\nu},\tag 5.6\\
\text{SM}\left(-\tfr14 \pi + 2\ell\pi, \ \ \tfr14\pi + 2\ell\pi\right)
& = -\i\sqrt{\frac{2}{\pi}}\,\Gamma(-\nu)\e^{-4\ell\pi\i\nu}\sin
\pi\nu.\tag 5.7\endalign
$$

To obtain (5.7) we need the asymptotics of $D_{-\nu-1}$, which are that
$$D_{-\nu-1}(\i z) \sim \cases
\e^{-\pi\i(\nu+1)/2}z^{-\nu-1}\exp(\tfr14z^2),\qquad &
\text{on} \ \arg z = -\frac 14\pi,\\
\displaystyle-\frac{\sqrt{2\pi}}{\Gamma(\nu+1)}\,\e^{-\pi
\i(\nu+2)/2}z^\nu\exp(-\tfr 14 z^2), \qquad & \text{on} \ \ \arg z = +\tfr
14\pi.\endcases 
$$

\subhead{6.  Monodromy Data for (3.1)}\endsubhead

Although one might expect the double turning-point case to be more
complicated than the simple case (and in some sense it is), yet in the
double turning-point case one can work out the monodromy data quite
explicitly, in terms of the coefficients of the monodromy equations,
even for a general form of equation.  It is this that leads to the
wealth of explicit connection formulae given, for example, in [11]. 
They are explicit because they are connecting directions where the
behaviour of the solution of the \pain \ equation leads to a double
turning-point in the isomonodromy equations.

In the present section, we show how this monodromy data can be
calculated.  To do this, we add to hypotheses H1-H3 in Theorem 1 the
following additional hypothesis.

\proclaim{H4}  Suppose that in H1 we can express $\widetilde F$ in the
form
$$\widetilde F(\eta,\xi) = F_1(\eta,\xi) + \xi^{-\gamma}F_2(\eta,\xi)$$
for some $\gamma > 0,$ where $F_1$ and $F_2$ are rational in $\eta.$ 
Suppose also that $F_0$ is a perfect square, so that
$F_1(\eta,\xi)/\{F_0^{1/2}(\eta)(\eta-\et)\}$ is rational in $\eta$
with partial fraction decomposition
$$\frac{F_1(\eta,\xi)}{F_0^{1/2}(\eta)(\eta-\et)} = \sum\limits^N_{i=1}
\frac{A_i}{\eta-s_i}, \qquad \text{with}\qquad s_1 = \et.\tag
6.1$$
Finally, suppose that, on and in a neighbourhood of $\ck, F_2/F_0^{1/2}$
is bounded uniformly in $\xi.$
\endproclaim

\remark{Remarks}
\item{1.}  The quantities $A_i,s_i$ will in general depend on $\xi,$
but we suppress that dependence.  We shall, however, assume that they
tend to finite limits as $|\xi|\to\infty.$
\medskip
\item{2.}  It is obvious that
$$A_1 = \frac{F_1(\et,\xi)}{F_0^{1/2}(\et)},\tag 6.2a$$
and we will set
$$B = \sum\limits^N_{i=1} A_i.\tag 6.2b$$
\medskip
\item{3.}  To compute the monodromy data, we need the relation between
$\z$ and $\eta$ for large $\xi.$  This is the content of the next
theorem.
\endremark

\proclaim{Theorem 3}  Under the hypothesis H4, and hypotheses H1-H3 of
Theorem 1, we have, for large $\xi$ and $\eta,$
$$\align
\z^2&-\alpha^2\log\z+\tfr14\alpha^2\log F_0(\et) +o(\xi^{-1})\\
&=2\int\limits^\eta_{\et} F_0^{1/2}(s)(s-\et)\thinspace\d s 
- \frac B\xi \log \eta
+ \frac 1 \xi \sum\limits^N_{i=2} A_i\log(\et-s_i).\tag 6.3\endalign
$$
\endproclaim
\demo{Proof}  From the definition of $\z$ and (3.16), we have
$$\tfr14\left\{2\z^2-2\alpha^2\log(2\z) + 2\alpha^2\log \alpha-\alpha^2 +
O(\alpha^4\z^{-2})\right\} = \int\limits^\eta_{\eta_2}F^{1/2}(s,\xi)
\thinspace \d s.\tag
6.4$$
\enddemo

In calculating the right-hand side, we will replace $F(\eta,\xi)$ by
$$\widehat F(\eta,\xi) =
F_0(\eta)(\eta-\et)^2-\frac{F_1(\eta,\xi)}{\xi},$$
i.e. we will ignore $F_2$.  This is justifiable because we will find
that even the $F_1$ term contributes only a term of size $O(\xi^{-1})$
which is all that we are interested in.  The term $F_2$ if we
included it, would similarly contribute a term of only
$O(\xi^{-1-\gamma}).$  Thus
$$\int\limits^\eta_{\eta_2}\widehat F^{1/2}(s,\xi)\thinspace\d s =
\left(\int\limits^{\eta^*}_{\eta_2} +
\int\limits^\eta_{\eta^*}\right)\widehat F^{1/2}(s,\xi)\thinspace \d s = I_1 +
I_2,\tag 6.5$$
say, where
$$\eta^* = \et + T\xi^{-1/2}$$
and $T$ is a large positive number to be specified more precisely later.
 In $I_1$ we make the change
$$s-\et = t\xi^{-1/2},$$
and then
$$I_1 = \frac 1\xi
\int\limits^T_{\{F_1(\et)/F_0(\et)\}^{1/2}}\{F_0(s)t^2-F_1(s)\}^{1/2}
\thinspace\d t.$$
Since $s-\et = O(\xi^{-1/2})$ and we are only concerned with evaluating
(6.5) correct to $O(\xi^{-1}),$ we can safely replace $s$ by $\et$ in
$I_1$ which, on integration using (3.16), gives
$$\align
I_1 = \frac{F_0^{1/2}(\et)}{4\xi}&\left\{2T^2- 
\frac{2F_1(\et)}{F_0(\et)}\log (2T) + \frac{F_1(\et)}{F_0(\et)} \log
\left(\frac{F_1(\et)}{F_0(\et)}\right) - \frac{F_1(\et)}{F_0(\et)}\right\}\\
&+O(\xi^{-1}T^{-2}) + o(\xi^{-1}).\tag 6.6\endalign
$$
Taking $T = -(F_1(\et)/F_0(\et))^{1/2},$ we can compute $I_1$
explicitly (with $s = \et)$ and so conclude from (3.5) that
$$\tfr12\i\pi\alpha^2 = \frac{F_0^{1/2}(\et)}{4\xi}\left[2\i\pi
\frac{F_1(\et)}{F_0(\et)}\right] + o(\xi^{-1}),$$
or
$$\alpha^2 = \frac{F_1(\et)}{\xi F^{1/2}_0(\et)} + o(\xi^{-1}).\tag
6.7$$
Also, by the binomial expansion, the integral $I_2$ in (6.5) is given by
$$\align I_2 = &\int^{\et}_{\eta^*}F^{1/2}_0(s)(s-\et)\left[1-\frac{F_1(s)}{2\xi
F_0(s)(s-\et)^2}\right] \d s \\ &\qquad+ O
\left(\int^\eta_{\eta^*}\left|\frac{F_1^2(s) \d
s}{\xi^2F^{3/2}_0(s)(s-\et)^3}\right|\right).\tag 6.8\endalign$$ According to hypothesis
H4, $F_1$ is bounded by $F_0^{1/2}$ and so the final term in this expression is
of size
$$O\left(\frac{1}{\xi^2} \int^\eta_{\eta^*}\frac{|\d
s|}{|F^{1/2}_0(s)||s-\et|^3}\right) = O \left(\frac{1}{\xi^2}
\frac{1}{|\eta^* -\et|^2}\right) = O(\xi^{-1}T^{-2}).$$ Using the proposed form
for $F_1(\eta,\xi)$ as given by (6.1), the second term in $I_2$ is equal to
$$\align
-\frac{1}{2\xi} \sum\limits^N_{i=1} &\int^\eta_{\eta^*}
\frac{A_i}{s-s_i} \d s = -\frac{B}{2\xi}\log \eta + O(\eta^{-1}\xi^{-1})
+ o(\xi^{-1})\\
&+ \frac{1}{2\xi} \log \prod\limits^N_{j=2}(\et-s_j)^{A_j}
+ \frac{F_1(\et)}{4\xi[F_0(\et)]^{1/2}}\log\left(\frac{T^2}{\xi}\right),\tag
6.9a\endalign
$$
whilst the first term may be written as
$$\int^\eta_{\eta^*}F^{1/2}_0(s)(s-\et)\thinspace\d s =
\int^\eta_{\et}F^{1/2}_0(s)(s-\et)\thinspace\d s - \tfr12
F^{1/2}_0(\et)T^2\xi^{-1} + o(\xi^{-1}).\tag 6.9b$$ Combining (6.5), (6.6),
(6.9a) and (6.9b) yields
$$\align
\int^\eta_{\eta^*}&F^{1/2}(s) \thinspace\d s = \frac{F_1(\et)}{4\xi
F^{1/2}_0(\et)}\left\{-2\log 2-\log \xi +
\log\left(\frac{F_1(\et)}{F_0(\et)}-1\right)\right\} 
-\frac{B}{2\xi} \log \eta \\
&+ \int^\eta_{\et} F^{1/2}_0(s)(s-\et)\thinspace\d s
+ \frac{1}{2\xi} \log
\prod\limits^N_{j=2}(\et-s_j)^{A_j} + O(\xi^{-1}T^{-2}) + o(\xi^{-1}),
\endalign
$$
and so, using (6.4) and (6.7) and making a choice of $T$ large, we see
that Theorem 3 is verified.

We have shown that linearly independent solutions of (3.8) are
$\nd(\pei\2xi\z)$ and $\1d(\ip\2xi \z)$ and the asymptotic behaviours of
these functions in various sectors have been noted in (5.1).  Using the
result of Theorem 3 and recalling that $\i\xi\alpha^2 = 2\nu + 1,$ it
follows that as $\xi, \z \to \infty,$
$$\align
&\z^{1/2}\e^{-\i\xi \z^2\!/2}\left(\pei\2xi \z\right)^\nu \sim
\e^{-\i\xi\Cal F(\eta)}\eta^{\i B/2}\\
&\times\left\{\e^{\i\xi\Cal
F(\et)}\xi^{\nu/2}\prod\limits^N_{j=2}(\et-s_j)^{-\i A_j/2}
F^{(2\nu+1)/8}_{00}2^{\nu/2}\e^{\i\pi\nu/4}\right\},\tag 6.10a\\
&\z^{1/2}\e^{\i\xi\z^2\!/2}\left(\pei\2xi \z\right)^{-\nu-1} \sim
\e^{\i\xi\Cal F(\eta)}\eta^{-\i B/2}\\
&\times\left\{\e^{-\i\xi\Cal
F(\et)}\xi^{-(1+\nu)/2}\prod\limits^N_{j=2}(\et - s_j)^{\i A_j/2}F_{00}^{-
(2\nu + 1)/8}2^{-(1+\nu)/2}\e^{-\i\pi(\nu+1)/4}\right\},\\
\tag 6.10b\endalign
$$
where $\Cal F(\eta) \equiv \int^\eta_0 F^{1/2}_0(s) (s-\et)\d s$ and
$F_{00}$ denotes the value $F_0(\et).$  These relations, together with
the Stokes multipliers (5.4)-(5.7) for the parabolic cylinder function,
enable us to write down the complete monodromy data for (3.1).  With $z
\equiv \pei\2xi \z,$ the Stokes multipliers relative to the solutions 
$$\e^{-\i\xi\Cal F(\eta)}\eta^{\i B/2}\quad \text{and}\quad
\e^{\i\xi\Cal F(\eta)}\eta^{-\i B/2}$$
are
\itemitem{i)}  from $\arg z = \frac 14 \pi + 2\ell \pi$ to $\frac 34
\pi + 2\ell \pi$ (where $\e^{-\i\xi\Cal F(\eta)}\eta^{\i B/2}$ is
the dominant solution)
$$\align
& -\frac{\sqrt{2\pi}}{\Gamma(-\nu)}\e^{\i \pi(4\ell + 1)\nu-2\i\xi\Cal
F(\et)}\xi^{-\nu-1/2}\\&\qquad\times\prod\limits^N_{j=2}(\et-s_j)^{\i
A_j}F_{00}^{-(2\nu + 1)/4} 2^{-\nu-1/2}\e^{-\i\pi(2\nu+1)/4};\tag 6.11a\endalign$$
\itemitem{ii)}  from $\arg z = \frac 34\pi + 2\ell \pi$ to $\frac 54\pi +
2\ell \pi$ ($\e^{\i\xi\Cal F(\eta)}\eta^{-\i B/2}$ dominant)
$$\align & \i\sqrt{\frac{2}{\pi}}\Gamma(-\nu)\e^{-\i\pi(4\ell+2)\nu+2\i\xi\Cal
F(\et)}\xi^{\nu+1/2}\\&\qquad\times\prod\limits^N_{j=2}(\et-s_j)^{-\i
A_j}F_{00}^{(2\nu+1)/4}2^{\nu+1/2}\e^{\i\pi(2\nu+1)/4}\sin\pi\nu;\tag 6.11b\endalign$$
\item\item{iii)}  from $\arg z = -\frac 34\pi + 2\ell \pi$ to $-\frac 14 \pi
+ 2\ell \pi$ ($\e^{-\i\xi\Cal F(\eta)}\eta^{\i B/2}$ dominant)
$$\align & \frac{\sqrt{2\pi}}{\Gamma(-\nu)}\e^{\i\pi(4\ell-1)\nu-2\i\xi\Cal
F(\et)}\xi^{-\nu-1/2}\\&\qquad\times\prod\limits^N_{j=2}(\et-s_j)^{\i
A_j}F_{00}^{-(2\nu+1)/4}2^{-\nu-1/2}\e^{-\i\pi(2\nu+1)/4};\tag 6.11c\endalign$$
\itemitem{iv)}  from $\arg z = -\frac 14\pi + 2\ell\pi$ to $\frac 14\pi +
2\ell \pi$ ($\e^{\i\xi\Cal F(\eta)}\eta^{-\i B/2}$ dominant)
$$\align & -\i\sqrt{\frac{2}{\pi}}\Gamma(-\nu)\e^{-4\i\pi\ell\nu+2\i\xi\Cal
F(\et)}\xi^{\nu+1/2}\\&\qquad\times\prod\limits^N_{j=2}(\et-s_j)^{-\i
A_j}F_{00}^{(2\nu + 1)/4}2^{\nu + 1/2}\e^{\i\pi(2\nu+1)/4}\sin \pi\nu.\tag
6.11d\endalign$$

We remark finally that, from Theorem 1 and (6.10), the asymptotic forms
of $\phi$ are 
$$F^{-1/4}\e^{-\i\xi\Cal F(\eta)}\eta^{\i B/2}\quad\text{and}\quad 
F^{-1/4}\e^{\i\xi\Cal F(\eta)}\eta^{-\i B/2}.\tag 6.12$$

\subhead{7.  Application to the \pain \ Equations}\endsubhead

Suppose that equation (3.1) arises after scaling from the monodromy
equation of some \pain \ equation.   (It is our contention that all such
monodromy equations reduce to the form (3.1) with simple or double
turning-points.)  In the preceding sections we have evaluated the
monodromy data with respect to the usual WKB solutions
$$F^{-1/4}\exp\left\{\pm \i \xi \int^\eta_{\et}F^{1/2}(t)\d t\right\}.$$  (These
solutions are given in terms of the variable $\eta$ but when expressed in
terms of the original variable $\lambda$ they are the usual WKB forms.) 
The theory of the \pain \ equations tells us that the monodromy data is
independent of $\xi$ provided that the $\lambda-$sector in which the
Stokes multipliers are being calculated remains fixed.  Since $\xi,
\lambda, \eta$ are inter-related (in the case of (2.2) $\eta =
\xi^{-1/3}\lambda),$ the condition that the $\lambda-$sector remains
fixed means that the $\eta-$sector changes with $\xi,$ or at least with
$\arg \xi,$ and also the turning-point $\et$ depends in general on $\xi.$
 Indeed, it may change from simple to double as $\arg \xi$ changes.  (In
the case of (2.2), this is a question of the behaviour of $M(\xi)$ as
$|\xi|\to\infty$ in a specific direction; $M(\xi) \to 0$ gives double
turning-points.)

Thus, the monodromy data depends on $\xi$ in various ways, but so long
as the $\lambda-$sector remains fixed, these various dependences must
cancel.  This leads, for example, to relations between $M(\xi)$ and
$\xi$ for large $|\xi|$, and so to statements about the possible
asymptotic behaviours of the \pain \ functions, which we will pursue in
later papers.

We now turn to examine the application of our method to the specific
case of PII (1.1) with $\beta = 0$ with the aim of using it to establish
Theorem B.  The relevant version of the generic equation (3.1) is
$$\align
\frac{\d^2\phi}{\d\eta} = \xi^2\phi&\left\{-(4\eta^2+1)^2+ \frac{8\i\eta}{\xi}
- \frac \i\xi (4\eta^2+1)\left(\eta-\frac{\i r}{2q\sqrt x}\right)^{-1} +
\frac{4r^2}{x^2} - \frac{4q^2}{x}-\frac{4q^4}{x^2}\right.\\
&\left.-\frac{2\i q^2\sqrt x}{\xi^2}\left(\eta-\frac{ \i r}{2q\sqrt x}\right)^{-1} +
\frac{3}{4\xi^2}\left(\eta-\frac{\i r}{2q\sqrt x}\right)^{-2}\right\},\tag
7.1\endalign 
$$
and this form follows directly from (2.2).  If either $x \to + \infty$ or
$x\to-\infty$, we will have a double turning-point $\et=\pm \frac 12
\i,$ and although we can choose, say, $\et = \frac 12 \i$ when we are
considering $x\to + \infty,$ the turning-point that we will have to use
when $x \to -\infty$ is then fixed.  Thus we have
$$F_0(\eta) = 16\left(\eta + \tfr12\i\right)^2\quad\text{if}\quad \et =
\tfr12\i, \quad F_0(\eta) = 16\left(\eta - \tfr12\i\right)^2\quad
\text{if}\quad \et = -\tfr12\i.\tag 7.2$$
In view of the behaviours of the solution of PII as $x\to\pm\infty$
(given by Theorem A and (1.2)), we will write
$$q = x^{-1/4}Q(\xi) = \xi^{-1/6}Q(\xi),\tag 7.3a$$
so that
$$r = \frac{\d q}{\d x} =\tfr32
\xi^{1/6}\left(Q'(\xi)-\frac{1}{6\xi}Q\right).\tag 7.3b$$ With these definitions
we will assume that there exists a sequence
$\xi_n\to \infty \e^{\i\theta}$ with
$$\frac{\i r}{2q\sqrt x} = \tfr34 \i \left(\frac{Q'}{Q} - \frac
{1}{6\xi}\right)\to \ell_1,\tag 7.4a$$
and 
$$\xi\left(\frac{4r^2}{x^2} - \frac{4q^4}{x^2} -
\frac{4q^2}{x}\right)\equiv 9\left[\left(Q'\right)^2-\frac{1}{3\xi} QQ' +
\frac{Q^2}{36\xi^2}\right] - \frac{4Q^4}{\xi} - 4 Q^2 \to \ell_2.\tag
7.4b$$
(This assumption is certainly justified if $\theta = 0, \frac 32\pi,$ and $q$
is the solution given by Theorem A.  Of course, $\ell_1$ and $\ell_2$
will depend on $\theta.$)  Then, with the notation of Section 6, as
$|\xi_n|\to\infty,$
$$F_1(\eta,\xi_n)\to 8\i\eta - \frac{\i(4\eta^2 + 1)}{\eta-\ell_1} +
\ell_2,\tag 7.5
$$
and, from (3.13) and (6.7), in the limit as $|\xi_n|\to\infty,$
$$2\nu + 1 - \i \frac{F_1(\et, \xi_n)}{F_0^{1/2}(\et)} \to 0,\tag
7.6a$$
so that
$$\nu + 1 \to \frac{\i\ell_2}{16\et}.\tag 7.6b$$
(Recall that we may have $\et = \frac 12\i$ or $\et = -\frac 12\i.$) 
Furthermore, $F_1/F^{1/2}$ has poles at $\eta =\pm \et$ and $\eta =
\ell_1$ and
$$\frac{F_1}{F^{1/2}} = -\frac{\i(2\nu + 1)}{\eta-\et} + \frac{\i(2\nu +
3)}{\eta + \et} - \frac{\i}{\eta-\ell_1},\tag 7.7$$
so that $s_2 = -\et, \quad s_3 = \ell_1, \quad A_2 = \i(2\nu + 3), \quad
A_3 = -\i$\quad and
$$B = \sum\limits^3_{j=1} A_j = \i.$$

We are now in a position to write down the respective monodromy data by
appealing to formulae (6.11).  However, rather than expressing the data
relative to the solutions (see (6.12))
$$\phi \sim F^{-1/4} \e^{-\i\xi\Cal F(\eta)}\eta^{\i
B/2}\quad\text{and}\quad \phi\sim F^{-1/4} \e^{\i\xi \Cal
F(\eta)}\eta^{-\i B/2},\tag 7.8$$
it is more convenient to use modified reference solutions.  We must use
$\psi$ rather than $\phi$ (see (2.1)), since it is in terms of $\psi$
and $\lambda$ that the monodromy data is independent of $\xi.$  As $\psi_2
= (\eta - l_1)^{1/2}\phi$ and since $B = \i,$ linearly independent
asymptotic solutions for $\psi_2$ are
$$\psi_2^{(1)} \sim \eta^{-1}\e^{-\i\xi\Cal F(\eta)}\quad\text{and}\quad
\psi_2^{(2)}\sim \e^{\i\xi\Cal F(\eta)}.$$
In order that $\Psi$ should satisfy (1.4), where the matrix has zero
trace, we need the component $\psi_1^{(1)} \sim\exp(-\i\xi\Cal F).$  It is
then immediate from (2.1) that $\psi_2^{(1)} \sim \frac 12 q\i
\xi^{-1/3}\eta^{-1}\exp(-\i\xi\Cal F),$ and so we choose to
establish monodromy data with respect to
$$\tfr12 q \i\xi^{-1/3}\eta^{-1}\e^{-\i\xi\Cal
F(\eta)}\quad\text{and}\quad \e^{\i\xi\Cal F(\eta)},$$
whence, from (6.11), for $\arg z \equiv \arg\left(\e^{\i\pi/4}\2xi \z\right) =
\frac 14\pi + 2\ell \pi$ to $\frac 34 \pi + 2\ell\pi,$ the Stokes
multiplier is
$$\align & -\frac{\i\sqrt{2\pi}}{\Gamma(-\nu)}\e^{(4\ell + 1)\i\pi\nu-2\i\xi\Cal
F(\et)}\xi^{-\nu-1/2}\left[\prod\limits^3_{j=2}(\et-s_j)^{\i A_j}\right]\\
&\qquad\times F_{00}^{-(\nu
+ 1/2)/2}2^{-\nu-3/2}\e^{-\i\pi(2\nu + 1)/4}q\xi^{-1/3},\tag 7.9a\endalign$$
whilst for the sector from $\arg z = \frac 34\pi + 2\ell \pi$ to $\frac
54 \pi + 2\ell \pi$ it is
$$\align &
\sqrt{\frac{2}{\pi}}\Gamma(-\nu)\e^{-(4\ell+2)\i\pi\nu+2\i\xi\Cal
F(\et)}\xi^{\nu + 1/2}\left[\prod\limits^3_{j=2}(\et-s_j)^{-\i A_j}\right]
\\ &\qquad\times F_{00}^{(\nu+1/2)/2}2^{\nu + 
3/2} \e^{\i\pi(2\nu+1)/4}q^{-1}\xi^{1/3}\sin \pi\nu.\tag 7.9b
\endalign
$$

\subhead{8.  Monodromy Data as $x \to + \infty$}\endsubhead

Here we take $\et = \frac 12\i,$ and the Stokes curves through $\et =
\frac 12\i$ are given by
$$\text{Re}\left\{\i\xi \int^\eta_{\i/2}\left(4\sigma^2 +
1\right)\d\sigma\right\} = \text{Re}\left[\i\xi\left(\tfr43 \eta^3 + \eta -
\tfr13 \i\right)\right] = 0,$$
which are asymptotic to the directions (with $\arg\xi = 0) \ \arg \eta =
\frac 13 j\pi,$ for integral $j.$  We shall choose the sector bounded by
$\arg \eta = 0$ and $\arg \eta = \frac 13\pi,$ which corresponds to the
$\lambda-$sector $0 \le \arg \lambda \le \frac 13\pi.$  This
$\lambda-$sector must then be the same when we consider $x\to -\infty.$

The asymptotics in Theorem A tell us that, as $x \to + \infty,$
$$\frac{\i r}{2q\sqrt x} \to -\tfr 12\i,$$
so that, in the notation of Section 7,
$$\ell_1 = -\tfr12\i, \ \ell_2 = 0, \ \nu = -1, \ s_2 = -\tfr 12\i, \
s_3 = -\tfr12\i, \ A_2 = \i, \ A_3=-\i, \ B = \i.$$
Also, from (6.10),
$$\Cal F(\et) = \tfr13\i, \quad F_{00} = -16.$$
Thus from (7.9) the Stokes multiplier for the relevant $z-$sector
$\left(\frac 14\pi \le \arg z \le \frac 34 \pi\right)$ is given by
$$\text{SM}_{\infty} = -a.\tag 8.1$$

\subhead{9.  Monodromy Data as $x \to -\infty$}\endsubhead

Since $\eta = x^{-1/2}\lambda$ and we now have $\arg x = \pi,$ the
requirement that the $\lambda-$sector be fixed now demands
$$-\tfr12 \pi \le \arg \eta \le -\tfr16 \pi.\tag 9.1$$
We assert that the relevant turning-point must now be $-\frac 12\i.$  For
if we suppose for contradiction that it is still $+\frac 12\i,$ then we
note that the Stokes curve from $\frac 12\i$ to $\infty \e^{-\i\pi/2}$
passes through $-\frac 12\i,$ since
$$\text{Re}\left\{\i\xi
\int^{-\i/2}_{\i/2}\left(4\sigma^2+1\right)\thinspace\d\sigma\right\} = 0.$$
(Recall that $\arg \xi = \frac 32 \pi.)$  Thus also the Stokes curve associated
with $-\frac 12\i$ and the sector (9.1) must pass through
$+\frac 12\i,$ and this is impossible since the real direction from
$-\frac 12\i$ is also a direction for which
$$\text{Re}\left\{\i\xi \int_{-\i/2}\left(4\sigma^2 +
1\right)\d\sigma\right\} = 0.$$
(Set $\sigma + \frac 12\i = \tau$ for small real $\tau.$)

The asymptotics in Theorem A tell us that, as $x\to -\infty,$
$$\frac{\i r}{2q\sqrt x} \sim -\tfr12 \cot \left(\tfr 23
\left|x\right|^{3/2} - \tfr34 d^2\log \left|x\right| + \gamma\right),$$
so that
$$\ell_1= - \tfr12 \underset{n \to \infty}\to\lim\left\{ \cot \left(\tfr 23
\left|x_n\right|^{3/2} - \tfr34d^2\log\left|x_n\right| + \gamma\right)\right\},$$
where the sequence $\{x_n\}$ (or $\{\xi_n\}$) has to be chosen so that the
limit exits.  Also
$$\ell_2 = 4\e^{3\pi \i/2}d^2, \quad \nu = - 1 + \tfr12 \i d^2,
\quad s_2 = \tfr 12\i,$$
$$s_3 = \ell_1, \quad A_2 = \i(2\nu + 3), \quad A_3 = -\i,$$
$$\Cal F(\et) = -\tfr 13\i, \qquad F_{00}= - 16.$$
Note also that since, from Theorem 3, $\arg \zeta = \frac 32 \arg \eta$
for large $|\eta|$, and since
$$\align
\arg z & = \tfr 14 \pi + \tfr 12\arg \xi + \arg \zeta\\
& = \tfr 14 \pi + \tfr 34 \arg x + \tfr 32 \arg \eta\\
& = \tfr 14 \pi + \tfr 32 \arg \lambda,\endalign
$$
we see that keeping the $\lambda-$sector fixed also fixes the $z-$sector
and so we have that the relevant $z-$sector is again $\frac 14 \pi \le
\arg z \le \frac 34 \pi.$  Thus from (7.9) the Stokes multiplier is

$$\align & \frac{\i\sqrt{2\pi}}{\Gamma(1-\frac 12\i d^2)} \e^{\pi
d^2\!/2}\e^{-2\xi/3}\xi^{(1-\i d^2)/2}(-\i)^{-(2\nu+3)}\left(-\tfr12\i -
\ell_1\right)\\ &\qquad\times\left(-16\right)^{(1-\i d^2)/4}  2^{-(1+
\i d^2)/2}\e^{{\i\pi}/{4} + \pi d^2\!/4} q\xi^{-1/3}.\tag 9.2\endalign $$
Since 
$$\tfr12\i + \ell_1 = -\tfr 12 \underset{n \to \infty}\to\lim\left\{
 \frac{\exp\{-\i(\frac 23|\xi_n| - \frac
12 d^2\log|\xi_n| + \gamma)\}}{\sin(\frac 23 |\xi_n| - \frac 12
d^2\log|\xi_n| + \gamma)}\right\}$$
we see that (9.2) reduces to
$$\text{SM}_{-\infty} = \frac{2\sqrt \pi}{d\Gamma(-\frac 12\i d^2)} \e^{-\pi
\i/4}\e^{-\i\gamma} 2^{-3\i d^2\!/2} \e^{-\pi d^2\!/4}.\tag 9.3$$
Since the Stokes multiplier must be independent of the $x-$direction,
comparison of (8.1) and (9.3) gives (1.3) and proves Theorem B.
\vfill\eject
\end